\documentclass[%
 reprint,
superscriptaddress,
 amsmath,amssymb,
 aps,
]{revtex4-2}

\usepackage{braket}
\usepackage{multirow}
\usepackage{graphicx}
\usepackage{dcolumn}
\usepackage{bm}
\usepackage[mathlines]{lineno}

\begin{document}

\preprint{APS/123-QED}

\title{Measurement of the transverse single-spin asymmetry for forward neutron production
in a wide $p_{\textrm{T}}$ range in polarized $p+p$ collisions at $\sqrt{s} = 510$ GeV}

\newcommand{\florence}{Department of Physics and Astronomy, University of Florence, Sesto Florentino, I-50019, Italy}
\newcommand{\florenceinfn}{INFN Section of Florence, Sesto Florentino, I-50019, Italy}
\newcommand{\riken}{RIKEN Nishina Center for Accelerator-Based Science, Wako, Saitama 351-0198, Japan}
\newcommand{\rikjrbrc}{RIKEN BNL Research Center, Brookhaven National Laboratory, Upton, 11973-5000, New York, USA}
\newcommand{\korea}{Korea University, Seoul 02841, Korea}
\newcommand{\sejong}{Sejong University, Seoul 05000, Korea}
\newcommand{\nagoyaieee}{Institute for Space-Earth Environmental Research, 
Nagoya University, Nagoya, 464-8602, Aichi, Japan}
\newcommand{\nagoyakmi}{Kobayashi-Maskawa Institute for the Origin of Particles 
and the Universe, Nagoya University, Nagoya, 464-8602, Aichi, Japan}
\newcommand{\shibaura}{Shibaura Institute of Technology, Minuma-ku, 337-8570, Saitama, Japan}
\newcommand{\bnl}{Brookhaven National Laboratory, Upton, 11973-5000, New York, USA}
\newcommand{\icrr}{Institute for Cosmic Ray Research, University of Tokyo, Kashiwa, 277-8582, Chiba, Japan}
\newcommand{\tokushima}{Tokushima University, 770-8051, Tokushima, Japan}
\newcommand{\jaea}{Advanced Science Research Center, Japan Atomic Energy 
Agency, 2-4 Shirakata Shirane, Tokai-mura, Naka-gun, Ibaraki-ken 319-1195, Japan}
\newcommand{\waseda}{RISE, Waseda University, Shinjuku, 162-0044, Tokyo, Japan}
\newcommand{\catania}{Department of Physics and Astronomy, University of Catania, Catania, I-95123, Italy}
\newcommand{\cataniainfn}{INFN Section of Catania, Catania, I-95123, Italy}
\newcommand{\csfnsm}{CSFNSM, Catania, I-95123, Italy}
\newcommand{\kek}{KEK, High Energy Accelerator Research Organization, Tsukuba, Ibaraki 305-0801, Japan}

\affiliation{\riken}
\affiliation{\florence}
\affiliation{\florenceinfn}
\affiliation{\rikjrbrc}
\affiliation{\korea}
\affiliation{\nagoyaieee}
\affiliation{\nagoyakmi}
\affiliation{\shibaura}
\affiliation{\sejong}
\affiliation{\bnl}
\affiliation{\kek}
\affiliation{\icrr}
\affiliation{\tokushima}
\affiliation{\jaea}
\affiliation{\waseda}
\affiliation{\catania}
\affiliation{\cataniainfn}
\affiliation{\csfnsm}

\author{M.~H.~Kim} \affiliation{\riken} 
\author{O.~Adriani} \affiliation{\florence} \affiliation{\florenceinfn}
\author{E.~Berti} \affiliation{\florence} \affiliation{\florenceinfn}
\author{L.~Bonechi} \affiliation{\florenceinfn}
\author{R.~D'Alessandro} \affiliation{\florence} \affiliation{\florenceinfn}
\author{Y.~Goto} \affiliation{\riken} \affiliation{\rikjrbrc}
\author{B.~Hong} \affiliation{\korea}
\author{Y.~Itow} \affiliation{\nagoyaieee} \affiliation{\nagoyakmi}
\author{K.~Kasahara} \affiliation{\shibaura}
\author{Y.~Kim} \affiliation{\sejong}
\author{J. H. Lee} \affiliation{\bnl}
\author{S. H. Lee} \affiliation{\sejong}
\author{T. Ljubicic} \affiliation{\bnl}
\author{H.~Menjo} \affiliation{\nagoyaieee}
\author{G. Mitsuka} \affiliation{\kek}
\author{I.~Nakagawa} \affiliation{\riken} \affiliation{\rikjrbrc}
\author{A.~Ogawa} \affiliation{\bnl}
\author{S.~Oh} \affiliation{\sejong}
\author{T.~Sako} \affiliation{\icrr}
\author{N.~Sakurai} \affiliation{\tokushima}
\author{K.~Sato} \affiliation{\nagoyaieee}
\author{R.~Seidl} \affiliation{\riken} \affiliation{\rikjrbrc}
\author{K.~Tanida} \affiliation{\jaea}
\author{S.~Torii} \affiliation{\waseda}
\author{A.~Tricomi} \affiliation{\catania} \affiliation{\cataniainfn} \affiliation{\csfnsm}


\collaboration{RHICf Collaboration}


\begin{abstract}
Transverse single-spin asymmetries $A_{\textrm{N}}$
of forward neutrons at pseudorapidities larger than
6 had only been studied in the transverse momentum range of $p_{\textrm{T}} < 0.4$ GeV/$c$.
The RHICf Collaboration has extended the previous measurements up to 1.0 GeV/$c$
in polarized $p+p$ collisions
at $\sqrt{s}~=~510$~GeV, using an electromagnetic calorimeter installed in the zero-degree area
of the STAR detector at the Relativistic Heavy Ion Collider.
The resulting $A_{\textrm{N}}$s increase in magnitude with $p_{\textrm{T}}$ in the high 
longitudinal momentum fraction $x_{\textrm{F}}$ range, but reach a plateau 
at lower $p_{\textrm{T}}$ for lower $x_{\textrm{F}}$.
For low transverse momenta the $A_{\textrm{N}}$s show little 
$x_{\textrm{F}}$ dependence and level off from intermediate values.
For higher transverse momenta the $A_{\textrm{N}}$s show also an indication to reach 
a plateau at increased magnitudes.
The results are consistent with previous measurements at lower collision energies, 
suggesting no $\sqrt{s}$ dependence of the neutron asymmetries.
A theoretical model based on the interference of 
$\pi$ and $a_1$ exchange between two protons
could partially reproduce the current results, however
an additional mechanism is 
necessary to describe the neutron $A_{\textrm{N}}$s over the whole
kinematic region measured.

\end{abstract}

\maketitle


\section{Introduction}
\label{sec:intro}

With discovery of a large transverse single-spin asymmetries ($A_{\textrm{N}}$)
for forward neutron production~\cite{ref:IP12} from the first polarized $p+p$ collisions
at a center of mass energy ($\sqrt{s}$) of 200 GeV at the
Relativistic Heavy Ion Collider (RHIC), 
the spin-dependent production mechanism of the forward
neutron has attracted great interest over ten years.
The discovery also inspired the PHENIX experiment to measure the neutron $A_{\textrm{N}}$s
at $\sqrt{s} =$ 62~GeV, 200~GeV, and 500~GeV~\cite{ref:PHENIX3}
at transverse momenta ($p_{\textrm{T}}$) less than 0.4 GeV/$c$ and indicated a
possible $p_{\textrm{T}}$ dependence of the neutron $A_{\textrm{N}}$.
The one-pion-exchange (OPE) model~\cite{ref:OPE1, ref:OPE2, ref:OPE3}, 
that successfully described the
unpolarized forward neutron production~\cite{ref:ISR}, 
introduced an interference between 
spin flip $\pi$ and spin nonflip $a_1$ exchange between the two protons.
This theoretical framework reproduced the PHENIX data reasonably well showing that the
neutron $A_{\textrm{N}}$s increased with increasing $p_{\textrm{T}}$ 
with little $\sqrt{s}$ dependence~\cite{ref:pi0_a1}.
Recently, the $A_{\textrm{N}}$s at $\sqrt{s} = 200$~GeV in Ref.~\cite{ref:PHENIX3}
were extracted as
function of longitudinal momentum fraction ($x_{\textrm{F}}$)
and $p_{\textrm{T}}$~\cite{ref:PHENIX_unfold}.
The results were consistent with the model calculations, but only relatively low 
transverse momenta were accessed.

The $A_{\textrm{N}}$ is defined by a left-right cross section asymmetry as
\begin{eqnarray}
A_{\textrm{N}} = \frac{d\sigma_{\textrm{left}} - d\sigma_{\textrm{right}}}
{d\sigma_{\textrm{left}} + d\sigma_{\textrm{right}}},
\label{eq:ANdef}
\end{eqnarray}
where $d\sigma_{\textrm{left} (\textrm{right})}$ is the particle production 
cross section in the
left (right) side of the beam polarization.
$A_{\textrm{N}}$s of forward particle production at pseudorapidities ($\eta$) 
larger than 6 at RHIC are especially important to study the production mechanism of the
particles in a region where perturbative quantum chromodynamics is not applicable.
Thus far the neutron $A_{\textrm{N}}$ has been studied
only in a narrow kinematic range in $p_{\textrm{T}} < 0.4$~GeV/$c$, 
measurements at higher $p_{\textrm{T}}>0.4$~GeV/$c$ have been awaited to study
the production mechanism of forward neutrons in more detail.
Here the RHIC forward (RHICf) Collaboration
has extended the kinematic range of the previous measurements up to
1.0 GeV/$c$ with one order of magnitude better position and $p_{\textrm{T}}$ resolutions
not only to explicitly explore the kinematic dependence of the neutron $A_{\textrm{N}}$
in a wide $p_{\textrm{T}}$ and $x_{\textrm{F}}$
ranges but also to study the $\sqrt{s}$ dependence
by comparing the results with those of PHENIX.

This paper is organized as follows.
The experimental setup and data taking of the RHICf experiment are presented in section 
\ref{sec:RHICf}.
The selection criteria for
good events and neutron candidates are explained in section 
\ref{sec:analysiscuts}.
Section \ref{sec:unfolding} describes the procedures of the background subtraction, unfolding,
and asymmetry calculation.
The results are discussed in section \ref{sec:results} and the paper is summarized in
section \ref{sec:summary}.

\section{The RHICf experiment}
\label{sec:RHICf}

\begin{figure}[b]
\centerline{%
\includegraphics[width=0.95\hsize]{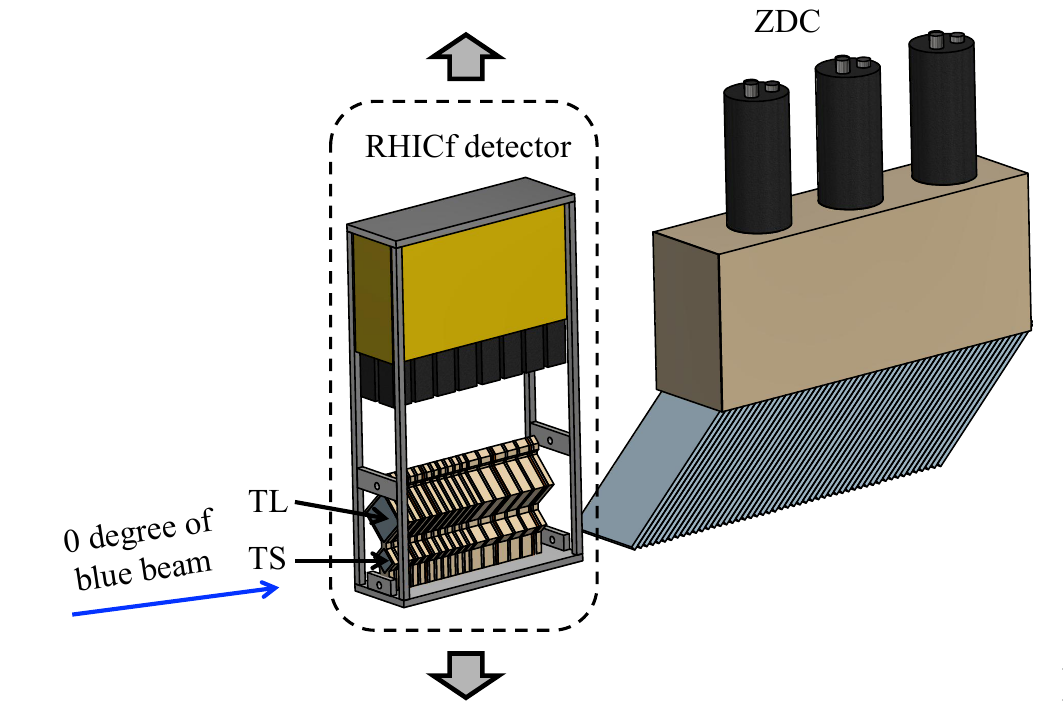}}
\caption{Setup of the RHICf experiment.
The data were taken by moving the RHICf detector to cover a wide $p_{\textrm{T}}$
range of $0.0 < p_{\textrm{T}} < 1.0$~GeV/$c$.}
\label{fig:setup}
\end{figure}
In June 2017,
the RHICf experiment measured forward neutral particles produced in
$\eta > 6$ from transversely polarized
$p+p$ collisions at $\sqrt{s} = 510$~GeV in the zero-degree area of the STAR
detector system at RHIC.
Figure~\ref{fig:setup} shows the experimental setup of the RHICf experiment.
STAR employs two
Zero-Degree Calorimeters (ZDCs)~\cite{ref:ZDC} located 18 m east and west,
from the nominal beam collision point.
The former LHCf Arm1 detector~\cite{ref:LHCfdet}, which will be called
RHICf detector~\cite{ref:RHICfdet} hereafter, was installed in front of the west ZDC.
A thin scintillator front counter (FC) was also positioned in front of the
RHICf detector to suppress charged hadron background.
The RHICf detector consists of two sampling calorimeters that have 
20~mm $\times$ 20~mm (small tower, TS) and 40~mm $\times$ 40~mm
(large tower, TL) effective areas, respectively.
Both are composed of 17 layers of tungsten absorbers with 1.6 nuclear interaction
lengths in total, 
16 layers of GSO scintillator plates, and 4 XY hodoscope layers covered by
1-mm-wide GSO bars.

RHICf used $90^{\circ}$-rotated 
transversely polarized beams (radially to the RHIC rings) instead of the usual vertically
polarized beams.
Neutrons with
a wide $p_{\textrm{T}}$ range of $0.0 < p_{\textrm{T}} < 1.0$~GeV/$c$ were
measured by moving the detector vertically.
We also requested large $\beta^*$ of 8~m for smaller angular beam divergence.
Under these conditions, the  luminosity was level of at
$10^{31}$~cm$^{-2}$s$^{-1}$.
See Ref.~\cite{ref:pi0paper} for more details on the experimental conditions.

\section{Event reconstruction and selection}
\label{sec:analysiscuts}

\begin{figure}[h]
\centerline{%
\includegraphics[width=0.8\hsize]{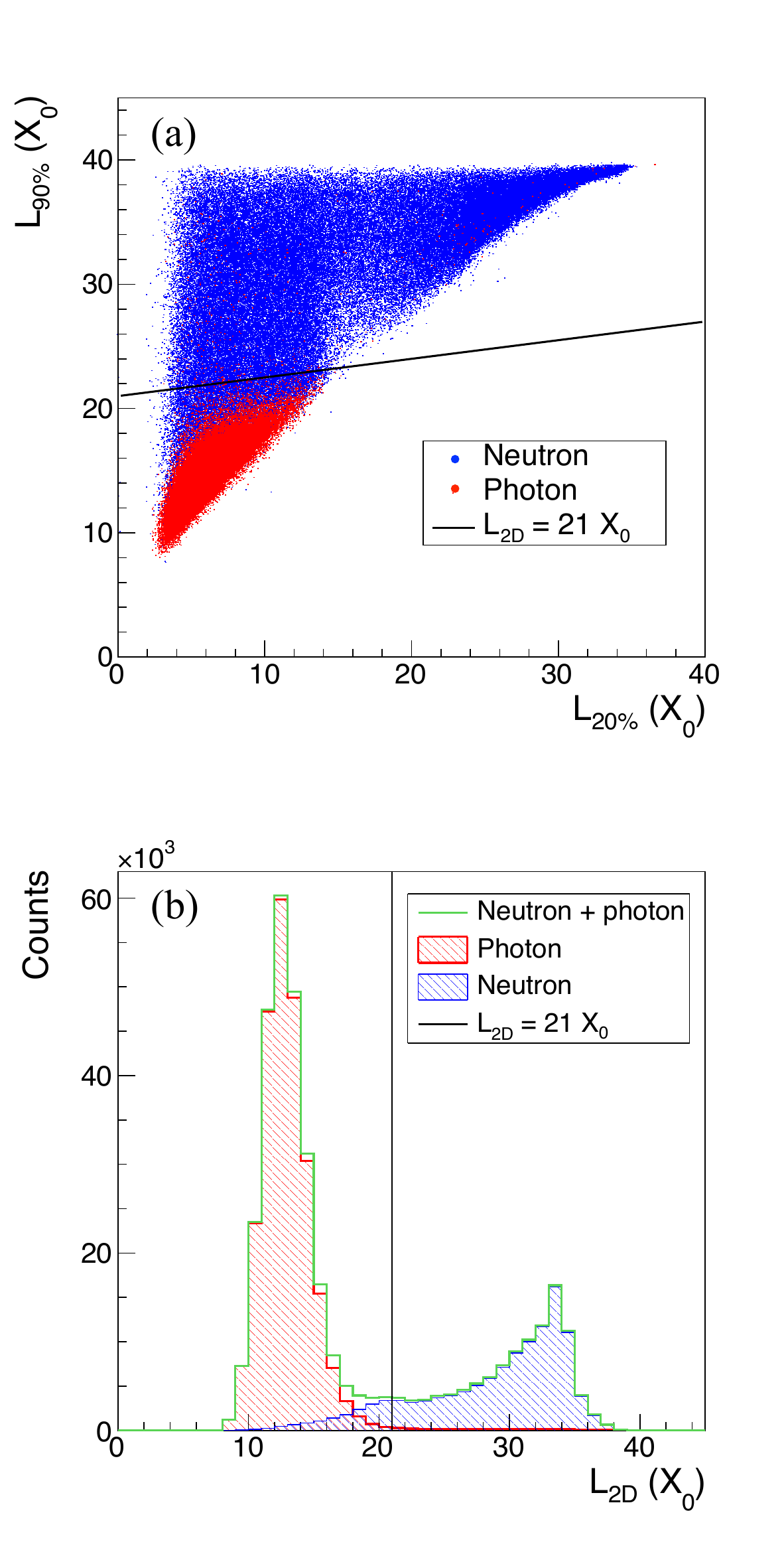}}
\caption{(a) $L_{90\%}$ versus $L_{20\%}$ and (b) $L_{2D}$ distributions of
neutron and photon events in the {\sc\small qgsjet ii-04} sample.
The black lines correspond to the threshold to select neutron candidate, which
is $L_{2D} = 21 X_{0}$.}
\label{fig:l2d}
\end{figure}

Before presenting the analysis selection criteria, 
the neutron and photon events are defined as follows.
A neutron event is defined as an event in which a neutron 
is produced by a collision and is directed toward the detector. When 
there is no neutron, a photon event is defined as an event in which 
at least one photon hits the detector.
The neutron events were mainly measured by the shower trigger that is activated when
the energy deposits of any three consecutive GSO plates are larger than 45~MeV.
Since the shower trigger is sensitive not only to the neutron events but also to
the photon events, the neutron candidates were identified by using the variable $L_{2D}$
defined by
\begin{eqnarray}
L_{2D} = L_{90\%}-0.15L_{20\%},
\label{eq:l2d}
\end{eqnarray} 
where $L_{x\%}$ is defined by the longitudinal depth for the measured 
integrated energy deposition in the GSO plates to reach $x\%$ of the total 
in units of the radiation length ($X_{0}$).
While neutrons mainly generate the hadronic showers in deeper layers of the
RHICf detector and do not necessarily deposit all their energy in the detector,
photons generate the electromagnetic shower in shallow layers
and deposit all their energy.
Figures~\ref{fig:l2d} (a) and (b) show the
$L_{90\%}$ versus $L_{20\%}$ and $L_{2D}$ distributions of the
neutron and photon events, respectively, 
in a Monte Carlo (MC) sample where the $p+p$ collisions
were generated by {\sc\small qgsjet ii-04}~\cite{ref:QGSJET}.
An event was identified as a neutron if the $L_{2D}$ was larger than $21~X_{0}$.
This threshold was optimized taking into account the neutron purity and efficiency
which were estimated by {\sc\small geant4}~\cite{ref:GEANT4} 
simulation with the {\sc\small qgsp$_{-}$bert 4.0} model.

Hit positions of the neutrons
were calculated by fitting the energy deposit distribution in the
GSO bars using a Lorentzian-based function.
One of the four hodoscope layers with the maximum energy deposition
was used for the position determination.
Energies of the neutrons were reconstructed using a relation between the energy deposit sum
of the GSO plates and the incident energy of neutrons obtained by 
{\sc\small geant4} simulations.
The position-dependent light collection efficiency and shower lateral leakage effect were also 
corrected in the simulation.
Although the energy range was different,
the above reconstructions were also applied for 
the previous analyses~\cite{ref:LHCfdet, ref:moredetails} 
that used the RHICf detector.
See Refs.~\cite{ref:moredetails2, ref:moredetails3, ref:RHICfdet} 
for more details on the reconstruction and correction procedures.

In order to study the detector performance for neutron reconstruction,
$10^5$ neutrons were generated to the center of the
detector in the {\sc\small geant4} simulation  
and their positions and energies were reconstructed in the same way as for the data.
For $200$~GeV neutrons, energy and position resolutions of the RHICf detector
were $1.1$~mm and $37\%$, respectively.
To improve the energy resolution, 
hadronic showers that developed deeper into the RHICf detector
were excluded by requiring $L_{90\%} <37~X_{0}$.
The condition improved the energy resolution of neutrons at, e.g., 200 GeV, from
$37\%$ to $30\%$.
The RHICf detector was located downstream of a RHIC dipole magnet, DX.
Neutron candidate hits were rejected if they overlapped with the shadow of the DX magnet, 
or their distance
to the detector edge was smaller than 2 mm because of the poor performance
in these regions.
In principle, only neutral particles can reach the detector from the collision point
because the DX magnet sweeps away charged particles.
However, the detector can detect charged particles when neutral hadrons 
hit the DX magnet and create a hadronic shower.
Events with ADC values of the FC larger than $25\%$ of the minimum ionizing
particle (MIP) peak position were excluded to suppress charged hadron background.

\section{Background subtraction and unfolding}
\label{sec:unfolding}

In the RHIC ring, the beam circulating clockwise is called ``blue beam" and
the one circulating counterclockwise ``yellow beam".
Since the RHICf detector was installed in the direction where the blue beam heads,
only the blue beam polarization was considered for the forward $A_{\textrm{N}}$ measurements.
On the other hand, when the backward $A_{\textrm{N}}$ was measured, only the
yellow beam polarization was taken into account.
Since RHICf used the beam polarization, which was normal to the direction that
the detector moved in Fig.~\ref{fig:setup},
the tower that was off-center of the beam measured only a narrow azimuthal
range of $\sigma_{\textrm{left (right)}}$ when the beam polarization was up (down).
In such case, the $A_{\textrm{N}}$ was defined by
\begin{eqnarray}
A_{\textrm{N}} = \frac{1}{PD_{\phi}}\Big(\frac{N^{\uparrow}-
RN^{\downarrow}}{N^{\uparrow}+RN^{\downarrow}}\Big),
\label{eq:luminosity}
\end{eqnarray}
where $P$ is the beam polarization, ranging from 0.54 to 0.61 for the blue beam and
from 0.53 to 0.61 for the yellow beam, and
$N^{\uparrow (\downarrow)}$ is the number
of neutrons detected when the beam polarization is up (down).
The beam polarization was measured by carbon target polarimeters~\cite{ref:p-Carbon}
and normalized by the absolute polarization measured 
by a hydrogen jet polarimeter~\cite{ref:H-jet}.
Systematic uncertainties of the blue and yellow beam polarizations
were $3.7\%$ and $3.4\%$, respectively.
$R$, estimated by the charged particle rates from the STAR's beam beam counter~\cite{ref:BBC}
and vertex position detector~\cite{ref:VPD}, is the ratio of luminosities with the polarization
of the blue beams up and down, and ranged from 0.958 to 0.995.
$D_{\phi}$ is a dilution factor estimated by
\begin{eqnarray}
D_{\phi} = \frac{1}{N}\sum_{i}\sin\phi_{i},
\end{eqnarray} 
where $\phi_i$ is the azimuthal angle of a neutron with respect to the beam polarization
in the $i$th event and $N$ is the number of total detected neutrons.
$D_{\phi}$ was used to compensate the dilution of $A_{\textrm{N}}$ originated from
a finite $\phi$ distribution of neutrons.
The average value of $D_{\phi}$ was 0.977.
If the neutron was measured by the tower on the beam center,
the azimuthal angles were divided into 8 equidistant bins and
the azimuthal modulation of the $A_{\textrm{N}}$ was measured by
\begin{eqnarray}
A_{\textrm{N}}(\phi) = \frac{1}{P} \Bigg(
\frac{\sqrt{N^{\uparrow}_{\phi}N^{\downarrow}_{\phi+\pi}} -
\sqrt{N^{\uparrow}_{\phi+\pi}N^{\downarrow}_{\phi}}}
{\sqrt{N^{\uparrow}_{\phi}N^{\downarrow}_{\phi+\pi}} +
\sqrt{N^{\uparrow}_{\phi+\pi}N^{\downarrow}_{\phi}}}
\Bigg),
\label{eq:square-root}
\end{eqnarray} 
where $N^{\uparrow(\downarrow)}_{\phi(\phi+\pi)}$ is the number of neutrons
detected in azimuthal angular bin $\phi(\phi+\pi)$ when the blue beam polarization
is up (down).
The $A_{\textrm{N}}$ was then calculated by fitting the azimuthal modulation with a 
sine function where magnitude and phase were left as free parameters.

\begin{figure}[t]
\centerline{%
\includegraphics[width=0.8\hsize]{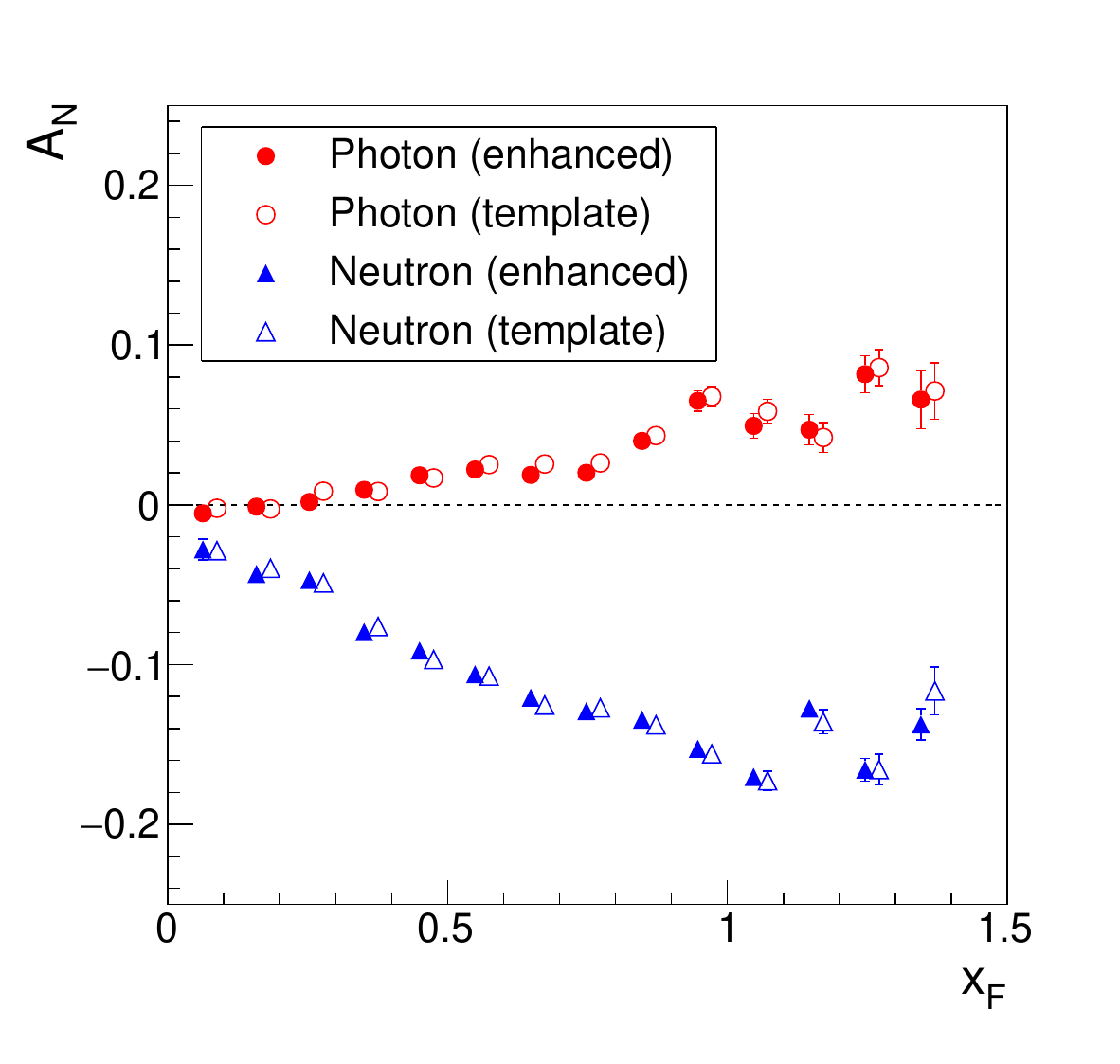}}
\caption{The neutron and photon $A_{\textrm{N}}$s calculated using the template fits
and enhanced samples.
Note that the $x_{\textrm{F}}$ is a reconstructed value that is not unfolded and
different $p_{\textrm{T}}$ bins were integrated.
The central $x_{\textrm{F}}$ values for the points from the template fit were shifted 
for better visibility.}
\label{fig:bgAN}
\end{figure}

In order to study the kinematic dependence of the neutron $A_{\textrm{N}}$,
$x_{\textrm{F}}$ and $p_{\textrm{T}}$ values were divided into equidistant intervals of
0.1 and 0.05 GeV/$c$, respectively.
Due to the finite position and energy resolutions of the detector, 
kinematic values of the neutrons 
were unfolded, but the background contaminations in the neutron candidates were 
subtracted first before unfolding.
Two background sources for the photon and charged hadron events were
considered.
\begin{figure}[b]
\centerline{%
\includegraphics[width=0.8\hsize]{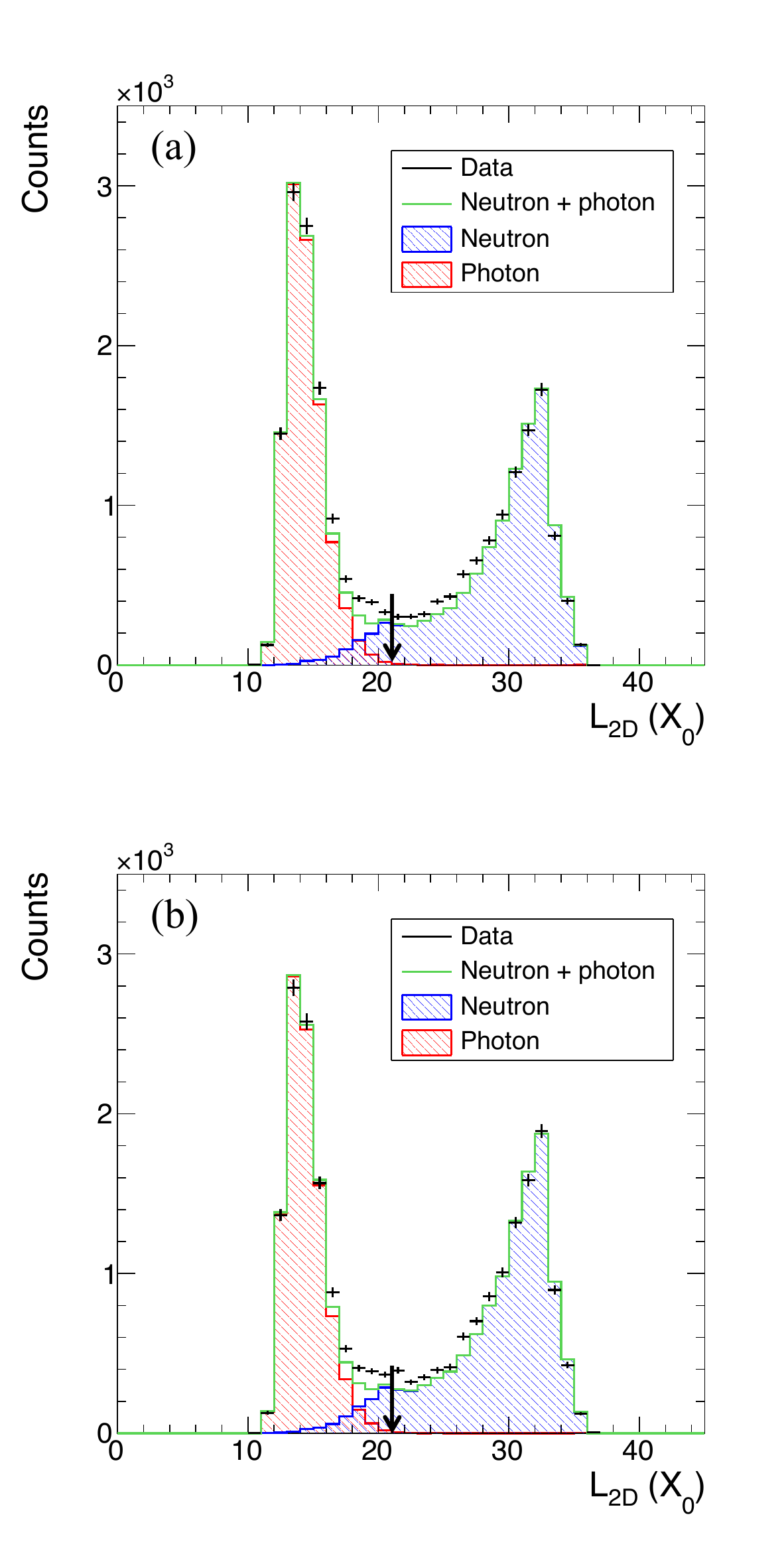}}
\caption{Template fit of the $L_{2D}$ distribution for the events where
the blue beam spin orientation is (a) up and (b) down.
The arrows show the threshold for selecting the neutron candidates, which is
$L_{2D} = 21 X_0$.
The kinematic range of the $L_{2D}$ distribution is $0.50<x_{\textrm{F}}<0.60$
and $0.30<p_{\textrm{T}}<0.35$ GeV/$c$.}
\label{fig:template}
\end{figure}
The contaminations in the two background event samples were subtracted for the
up and down polarization cases separately:
\begin{eqnarray}
N_{\textrm{neu}}^{\uparrow} = N_{\textrm{trig}}^{\uparrow}
- N_{\textrm{pho}}^{\uparrow} - N_{\textrm{cha}}^{\uparrow} \\
N_{\textrm{neu}}^{\downarrow} = N_{\textrm{trig}}^{\downarrow}
- N_{\textrm{pho}}^{\downarrow} - N_{\textrm{cha}}^{\downarrow},
\end{eqnarray}
where $N_{\textrm{trig}}^{\uparrow (\downarrow)}$,
$N_{\textrm{neu}}^{\uparrow (\downarrow)}$, $N_{\textrm{pho}}^{\uparrow (\downarrow)}$,
and $N_{\textrm{cha}}^{\uparrow (\downarrow)}$ are the number of triggered, neutron,
photon, and charged hadron events, respectively, when the blue beam polarization
is up (down).
The charged hadron events are defined as at least one charged hadron hits the detector
when there is no neutron produced by the collision, that heads towards the detector.
In order to estimate the $N_{\textrm{pho}}^{\uparrow} $ and $N_{\textrm{pho}}^{\downarrow}$,
we performed a template fit of the $L_{2D}$ distribution by scaling the neutron
and photon events of the same kinematic bin in the {\sc\small qgsjet ii-04} sample separately.
Figure~\ref{fig:template} shows an example of the template fit for one kinematic bin.
The down-to-up ratios of the neutron and photon events,
$N^{\downarrow}_{\textrm{neu}}/N^{\uparrow}_{\textrm{neu}}$ and 
$N^{\downarrow}_{\textrm{pho}}/N^{\uparrow}_{\textrm{pho}}$, in Fig.~\ref{fig:template}
estimated by the scaled templates are
$1.077\pm0.014$ and $0.920\pm0.012$, which is consistent with the sign of the 
previously measured neutron~\cite{ref:IP12, ref:PHENIX3} 
and $\pi^0$ asymmetries~\cite{ref:pi0paper}.
Figure~\ref{fig:bgAN} shows the $A_{\textrm{N}}$s of the neutron and photon events
calculated using the template fits and enhanced samples before unfolding.
The neutron and photon enhanced samples were selected by applying $L_{2D} > 21 X_{0}$
and $L_{90\%} < 18 X_{0}$~\cite{ref:pi0paper}, respectively.
Consistencies between the two $A_{\textrm{N}}$s calculated by the above two methods prove that
the numbers of neutrons and photons were correctly estimated by the template fit.
The photon contamination estimated by the template fit, which was less than $0.7\%$ of the
total neutron candidate sample was subtracted.
In Fig.\ref{fig:template}, The larger $L_{2D}$ values of data in $15<L_{2D}<21 X_{0}$ indicate that
the photon energy distribution of data is higher than that of MC because photons with higher
energy generally deposit energy over a larger longitudinal region, making the $L_{2D}$ value
larger than for lower-energy photons.
To study the effect of the discrepancies, the photon event template of the
$i$th $x_{\textrm{F}}$ bin 
was replaced by the one of the $i+1$th $x_{\textrm{F}}$ bin.
The template fit was improved, but the
$A_{\textrm{N}}$ difference between the two template fits after unfolding
was negligible, which was less than 0.0007.
We concluded that the effect of the discrepancies was negligible, 
thereby we did not consider the systematic uncertainty of the template fit.

\begin{figure}[t]
\centerline{%
\includegraphics[width=0.8\hsize]{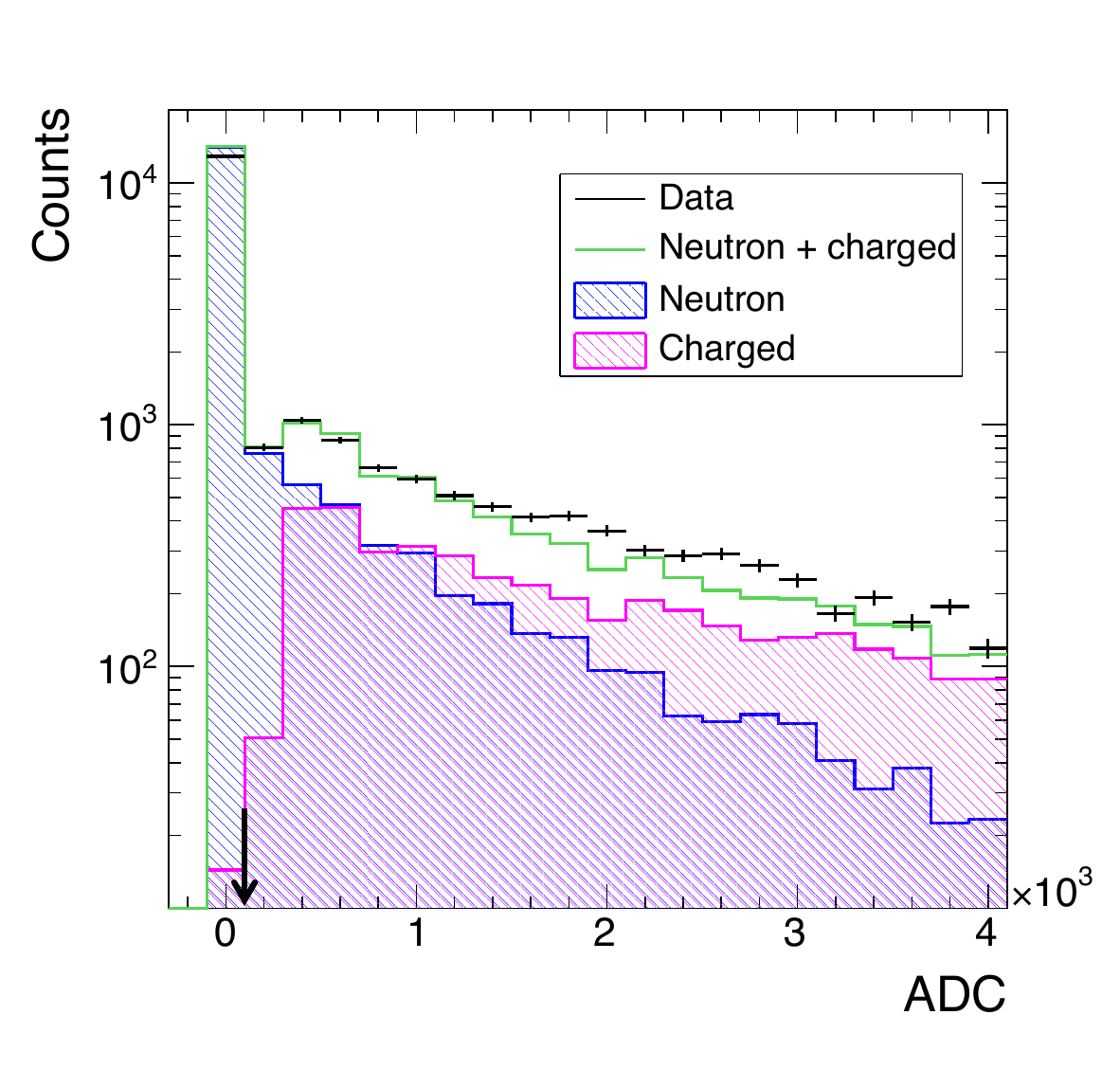}}
\caption{Template fit of the ADC distribution of FC when the
blue beam is polarized up.
The arrow shows the threshold to suppress the charged hadron background,
which is ADC$>$0.25MIP.
The kinematic range of the ADC distribution is $0.10<x_{\textrm{F}}<0.20$ and
$0.05<p_{\textrm{T}}<0.10$ GeV/$c$.}
\label{fig:FCadc}
\end{figure}

Another template fit was performed to the ADC distribution of the FC to estimate the 
$N_{\textrm{cha}}^{\uparrow}$ and $N_{\textrm{cha}}^{\downarrow}$ by
scaling the neutron and charged hadron event templates of the same kinematic bin 
in the {\sc\small qgsjet ii-04} sample separately.
Fig. \ref{fig:FCadc} shows an example of the template fit.
The average contamination of charged hadron events in the neutron candidate sample,
which was selected by applying $L_{2D}>21 X_0$, was
$0.2\%$, which was subtracted from the up and down polarization events separately.
Since the template fit of the ADC distribution was an independent process of the
one performed to the $L_{2D}$ distribution, the two following cases were considered
to study the systematic uncertainty in the charged hadron subtraction process:
every charged hadron event (1) had at least one photon and (2) did not have any photon.
In the case of (1), 
only the photon contamination
was subtracted because the charged hadron contamination was less
than the photon.
In the case of (2), the two contaminations were subtracted respectively.
The difference between the two cases was negligible on the $A_{\textrm{N}}$s,
being less than 0.0004.
Therefore, we also did not assign a systematic uncertainty to the process of the
charged hadron subtraction.
According to {\sc\small qgsjet ii-04}, 
the neutron  candidate sample was composed to $95.0\%$ of neutrons, 
$3.5\%$ $\Lambda$s, and $1.5\%$ neutral kaons, after background subtraction.

The kinematic values of the neutrons, $x_{\textrm{F}}$, $p_{\textrm{T}}$, and $\phi$, 
were unfolded
using the Bayesian unfolding method~\cite{ref:Bayes} 
as implemented in the
RooUnfold~\cite{ref:RooUnfold} package of {\sc\small root}~\cite{ref:root}.
For prior, a MC sample where 
the neutrons from 0 to 255 GeV were uniformly generated on 
the detector was used to avoid any bias from the particular particle productions.
The iterative procedure was stopped when
the $\chi^2$ change between two outputs of consecutive
iterations became smaller than 1.
The variation of $A_{\textrm{N}}$ by uncertainties of the unfolded data points 
was considered as one of the systematic uncertainties.
This uncertainty is the dominating systematic uncertainty.
We generated finite asymmetries by assigning up and down spin patterns in the 
{\sc\small qgsjet~ii-04} sample and confirmed that the unfolded spectra
reproduced the input $\braket{x_{\textrm{F}}}$, $\braket{p_{\textrm{T}}}$, and
 $A_{\textrm{N}}$ well within the total uncertainty that included the statistical and
 systematic uncertainties.
 The differences between the reconstructed and input $\braket{x_{\textrm{F}}}$ and
 $\braket{p_{\textrm{T}}}$ were less than 0.04 and 0.02~GeV/$c$, respectively.
Besides the systematic uncertainty of the unfolding process, the uncertainty of
the beam center calculation was also considered.
The beam center was measured by two methods~\cite{ref:pi0paper} and
half of the $A_{\textrm{N}}$ 
difference between the two methods was assigned as systematic uncertainty.

\section{Results}
\label{sec:results}

\begin{figure}[t]
\centerline{%
\includegraphics[width=1.1\hsize]{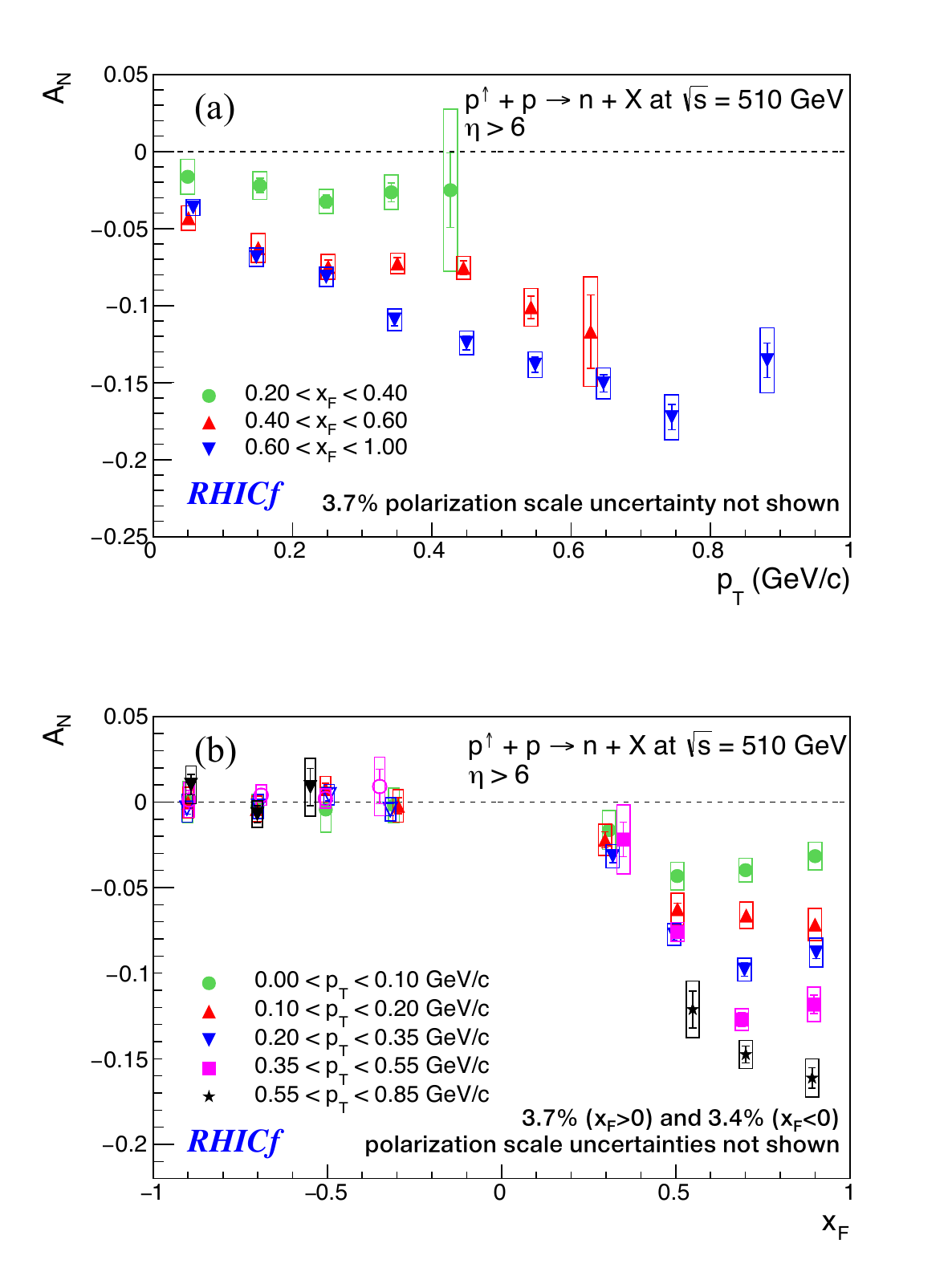}}
\caption{$A_{\textrm{N}}$ for forward neutron production as function of
(a) $p_{\textrm{T}}$ and (b) $x_{\textrm{F}}$.
Error bars correspond to the statistical uncertainties, and the boxes represent the
total systematic uncertainties.}
\label{fig:result}
\end{figure}
\begin{figure}[t]
\centerline{%
\includegraphics[width=1.1\hsize]{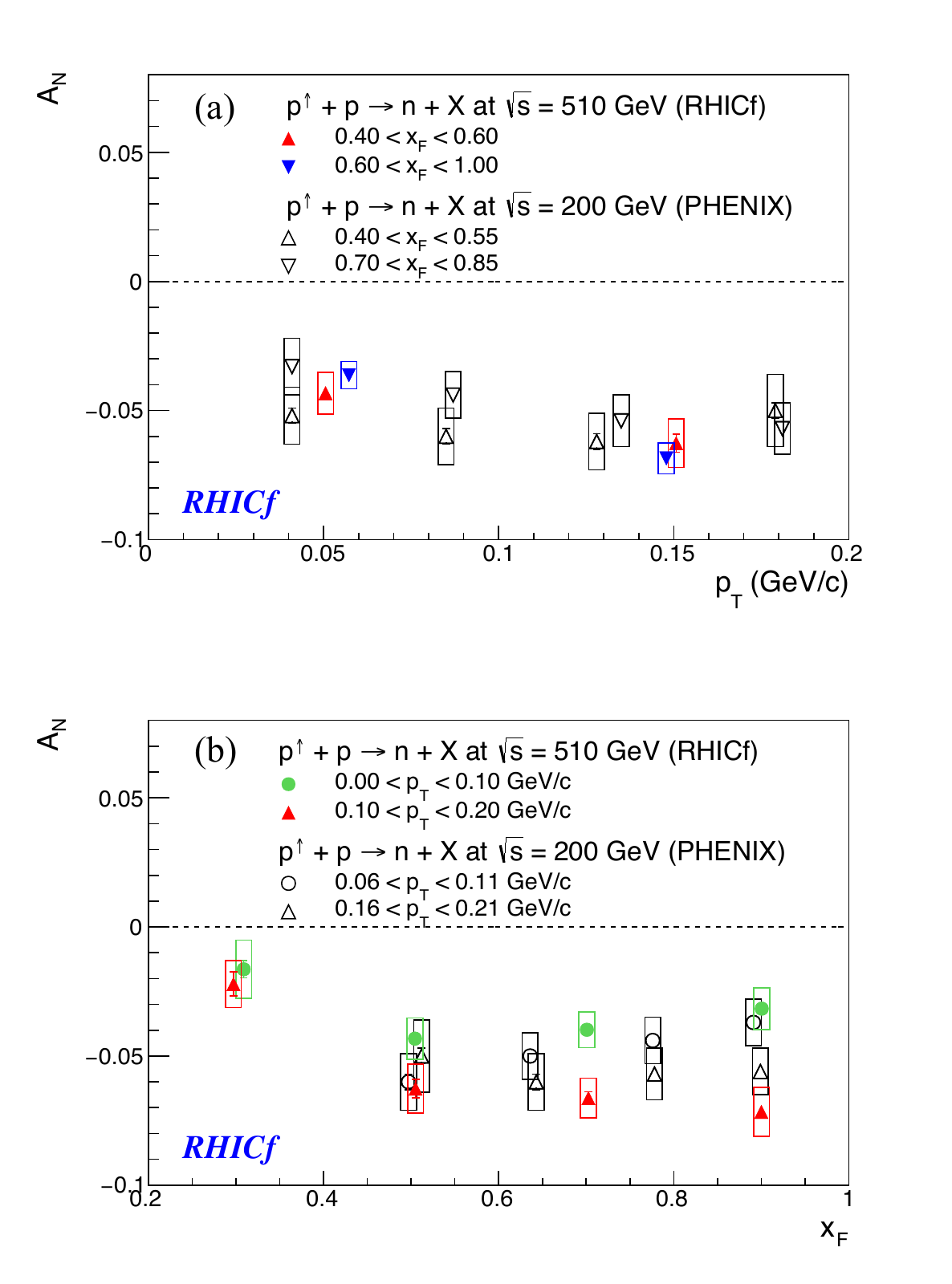}}
\caption{Comparison of the RHICf results with those of PHENIX 
as function of (a) $p_{\textrm{T}}$ and (b) $x_{\textrm{F}}$.}
\label{fig:comp}
\end{figure}
Figure~\ref{fig:result}, Table~\ref{tb:forward}, and Table ~\ref{tb:backward}
summarize the $A_{\textrm{N}}$s for forward neutron production
as function of $\braket{x_{\textrm{F}}}$ and $\braket{p_{\textrm{T}}}$ measured by the
RHICf experiment.
Figure~\ref{fig:result} (a) shows the neutron $A_{\textrm{N}}$s as a function of 
$p_{\textrm{T}}$ in three different $x_{\textrm{F}}$ ranges.
In the low $x_{\textrm{F}}$ range, the neutron $A_{\textrm{N}}$ reaches a plateau
at low $p_{\textrm{T}}$.
In the high $x_{\textrm{F}}$ range, the plateau does not seem to be reached yet
while the absolute value of the $A_{\textrm{N}}$ explicitly increases in magnitude with
$p_{\textrm{T}}$.
Figure~\ref{fig:result} (b) shows the $A_{\textrm{N}}$s as a function of $x_{\textrm{F}}$ in
five different $p_{\textrm{T}}$ ranges.
The backward $A_{\textrm{N}}$s are all consistent with zero.
In the low $p_{\textrm{T}}$ range $< 0.20$~GeV/$c$, 
the forward $A_{\textrm{N}}$ reaches
a plateau of low $A_{\textrm{N}}$ at low $x_{\textrm{F}}$ (about 0.5) with little 
$x_{\textrm{F}}$ dependence.
In the high $p_{\textrm{T}}$ range $>0.20$~GeV/$c$, 
the asymmetries appear to be leveling off at higher $x_{\textrm{F}}$
(about 0.7), showing a clear $x_{\textrm{F}}$ dependence.
The $x_{\textrm{F}}$ dependence in the high $p_{\textrm{T}}$ range was observed for the
first time by the RHICf experiment.
Figure~\ref{fig:comp} (a) shows the comparison between the RHICf and PHENIX data
as a function of $p_{\textrm{T}}$.
In the range of low $p_{\textrm{T}}$ $< 0.2$ GeV/$c$ and $x_{\textrm{F}} > 0.4$
that is overlapping with the PHENIX data at $\sqrt{s}$= 200 GeV, 
the asymmetries are consistent with those by RHICf at $\sqrt{s}$ = 510~GeV.
Figure~\ref{fig:comp} (b) shows the comparison between the two experiments as a function of 
$x_{\textrm{F}}$.
In the low $p_{\textrm{T}}$ range
that PHENIX covers at $\sqrt{s}$=200 GeV, the asymmetries are again consistent at both energies 
and show a flat $x_{\textrm{F}}$ dependence. 
Figures~\ref{fig:comp}~(a) and (b) suggest that there is no or only a weak $\sqrt{s}$ dependence.

\begin{figure}[t]
\centerline{%
\includegraphics[width=1.1\hsize]{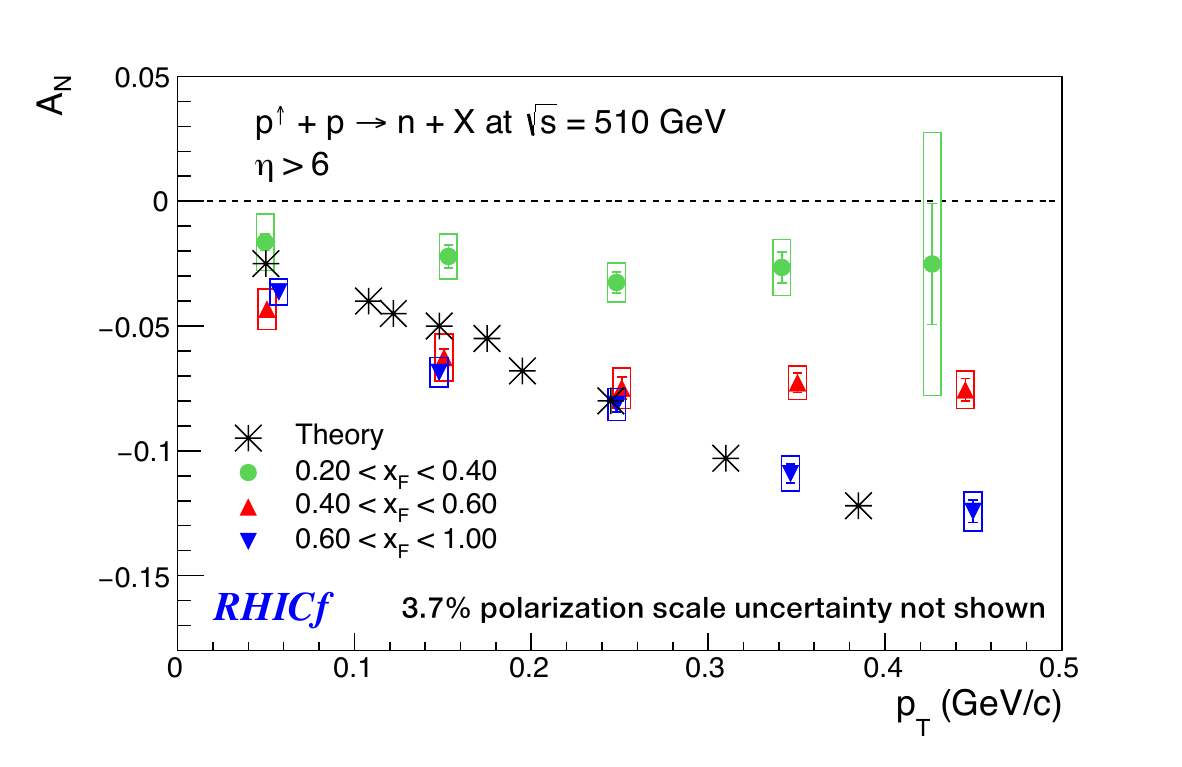}}
\caption{Comparison of the RHICf results with the theoretical calculations.}
\label{fig:comp2}
\end{figure}

The RHICf data is also compared to model calculations~\cite{ref:pi0_a1} 
based on the $\pi$ and $a_1$ exchange, as shown
in Fig.~\ref{fig:comp2}.
The model did not predict the $x_{\textrm{F}}$ dependence of the neutron $A_{\textrm{N}}$.
In the high $x_{\textrm{F}}$ range,
the $A_{\textrm{N}}$s are mostly consistent with the model calculations. 
However, the model does not reproduce the $A_{\textrm{N}}$s in the low $x_{\textrm{F}}$ range
where the asymmetries are significantly smaller.
This may be because fragmentation is expected to dominate neutron production at 
low $x_{\textrm{F}}$ over Reggeon exchange.

\begin{table*}[ht]
\begin{center}
\begin{tabular}{c  c  c  c  c  c  c}
\hline \hline
\multirow{2}*{$\langle x_{\textrm{F}}\rangle$} & 
\multirow{2}*{$\langle p_{\textrm{T}}\rangle$ (GeV/$c$)} & 
\multirow{2}*{$A_{\textrm{N}}$} &
\multirow{2}*{Statistical uncertainty} & \multicolumn{3}{c}{Systematic uncertainty}\\
\cline{5-7} & & & & Total & Beam center & Unfolding\\
\hline
$0.30$ & $0.04$ & $-0.0164$ & $0.0033$ & $0.0112$ & $0.0003$ & $0.0112$\\
$0.29$ & $0.15$ & $-0.0221$ & $0.0046$ & $0.0090$ & $0.0006$ & $0.0089$\\
$0.31$ & $0.24$ & $-0.0325$ & $0.0041$ & $0.0077$ & $0.0002$ & $0.0077$\\
$0.34$ & $0.34$ & $-0.0262$ & $0.0062$ & $0.0111$ & $0.0008$ & $0.0109$\\
$0.34$ & $0.42$ & $-0.0251$ & $0.0242$ & $0.0526$ & $0.0361$ & $0.0383$\\[1ex]

$0.50$ & $0.05$ & $-0.0432$ & $0.0022$ & $0.0077$ & $0.0006$ & $0.0077$\\
$0.50$ & $0.15$ & $-0.0626$ & $0.0034$ & $0.0087$ & $0.0047$ & $0.0073$\\
$0.49$ & $0.25$ & $-0.0748$ & $0.0044$ & $0.0079$ & $0.0025$ & $0.0075$\\
$0.49$ & $0.35$ & $-0.0727$ & $0.0039$ & $0.0065$ & $0.0003$ & $0.0065$\\
$0.50$ & $0.44$ & $-0.0755$ & $0.0044$ & $0.0071$ & $0.0004$ & $0.0071$\\
$0.54$ & $0.54$ & $-0.1011$ & $0.0072$ & $0.0118$ & $0.0004$ & $0.0118$\\
$0.54$ & $0.62$ & $-0.1169$ & $0.0237$ & $0.0354$ & $0.0117$ & $0.0334$\\[1ex]

$0.80$ & $0.05$ & $-0.0363$ & $0.0014$ & $0.0051$ & $0.0004$ & $0.0051$\\
$0.80$ & $0.15$ & $-0.0685$ & $0.0015$ & $0.0057$ & $0.0012$ & $0.0056$\\
$0.79$ & $0.24$ & $-0.0814$ & $0.0030$ & $0.0060$ & $0.0001$ & $0.0060$\\
$0.80$ & $0.34$ & $-0.1091$ & $0.0038$ & $0.0067$ & $0.0003$ & $0.0067$\\
$0.74$ & $0.44$ & $-0.1242$ & $0.0044$ & $0.0071$ & $0.0006$ & $0.0071$\\
$0.75$ & $0.54$ & $-0.1383$ & $0.0048$ & $0.0078$ & $0.0003$ & $0.0078$\\
$0.76$ & $0.64$ & $-0.1504$ & $0.0056$ & $0.0095$ & $0.0020$ & $0.0093$\\
$0.81$ & $0.74$ & $-0.1724$ & $0.0081$ & $0.0137$ & $0.0018$ & $0.0136$\\
$0.88$ & $0.88$ & $-0.1355$ & $0.0111$ & $0.0200$ & $0.0057$ & $0.0192$\\
\hline\hline
\end{tabular}
\caption{$A_{\textrm{N}}$s for forward neutron production as function of 
$\braket{x_{\textrm{F}}}$ and $\braket{p_{\textrm{T}}}$.
$A_{\textrm{N}}$ and kinematic values from Fig. \ref{fig:result} (a) are listed.} 
\label{tb:forward}
\end{center}
\end{table*}

\begin{table*}[ht]
\begin{center}
\begin{tabular}{c  c  c  c  c  c  c}
\hline \hline
\multirow{2}*{$\langle x_{\textrm{F}}\rangle$} & 
\multirow{2}*{$\langle p_{\textrm{T}}\rangle$ (GeV/$c$)} & 
\multirow{2}*{$A_{\textrm{N}}$} &
\multirow{2}*{Statistical uncertainty} & \multicolumn{3}{c}{Systematic uncertainty}\\
\cline{5-7} & & & & Total & Beam center & Unfolding\\
\hline
$-0.89$ & $0.67$ & $0.0100$ & $0.0059$ & $0.0107$ & $0.0017$ & $0.0106$\\
$-0.70$ & $0.64$ & $-0.0071$ & $0.0049$ & $0.0075$ & $0.0007$ & $0.0075$\\
$-0.54$ & $0.58$ & $0.0087$ & $0.0108$ & $0.0168$ & $0.0051$ & $0.0160$\\[1ex]

$-0.89$ & $0.43$ & $0.0011$ & $0.0054$ & $0.0099$ & $0.0013$ & $0.0098$\\
$-0.69$ & $0.44$ & $0.0039$ & $0.0038$ & $0.0059$ & $0.0003$ & $0.0059$\\
$-0.50$ & $0.43$ & $0.0019$ & $0.0032$ & $0.0053$ & $0.0004$ & $0.0052$\\
$-0.34$ & $0.38$ & $0.0090$ & $0.0100$ & $0.0171$ & $0.0012$ & $0.0170$\\[1ex]

$-0.90$ & $0.27$ & $-0.0035$ & $0.0035$ & $0.0081$ & $0.0021$ & $0.0078$\\
$-0.69$ & $0.26$ & $-0.0025$ & $0.0039$ & $0.0066$ & $0.0014$ & $0.0064$\\
$-0.49$ & $0.27$ & $0.0042$ & $0.0035$ & $0.0059$ & $0.0005$ & $0.0059$\\
$-0.31$ & $0.26$ & $-0.0043$ & $0.0036$ & $0.0067$ & $0.0003$ & $0.0067$\\[1ex]

$-0.90$ & $0.15$ & $-0.0002$ & $0.0020$ & $0.0090$ & $0.0004$ & $0.0090$\\
$-0.70$ & $0.14$ & $-0.0040$ & $0.0023$ & $0.0075$ & $0.0012$ & $0.0074$\\
$-0.50$ & $0.15$ & $0.0074$ & $0.0034$ & $0.0074$ & $0.0000$ & $0.0074$\\
$-0.29$ & $0.15$ & $-0.0022$ & $0.0047$ & $0.0090$ & $0.0005$ & $0.0090$\\[1ex]

$-0.90$ & $0.06$ & $0.0000$ & $0.0019$ & $0.0121$ & $0.0000$ & $0.0121$\\
$-0.70$ & $0.05$ & $-0.0025$ & $0.0019$ & $0.0069$ & $0.0002$ & $0.0069$\\
$-0.50$ & $0.05$ & $-0.0043$ & $0.0022$ & $0.0126$ & $0.0000$ & $0.0126$\\
$-0.30$ & $0.04$ & $-0.0019$ & $0.0033$ & $0.0100$ & $0.0000$ & $0.0100$\\[1ex]

$0.30$ & $0.49$ & $-0.0164$ & $0.0033$ & $0.0112$ & $0.0003$ & $0.0112$\\
$0.50$ & $0.50$ & $-0.0432$ & $0.0022$ & $0.0077$ & $0.0006$ & $0.0077$\\
$0.70$ & $0.54$ & $-0.0398$ & $0.0019$ & $0.0067$ & $0.0004$ & $0.0067$\\
$0.90$ & $0.60$ & $-0.0316$ & $0.0019$ & $0.0078$ & $0.0011$ & $0.0077$\\[1ex]

$0.29$ & $0.15$ & $-0.0221$ & $0.0046$ & $0.0090$ & $0.0006$ & $0.0089$\\
$0.50$ & $0.15$ & $-0.0626$ & $0.0034$ & $0.0087$ & $0.0047$ & $0.0073$\\
$0.70$ & $0.14$ & $-0.0662$ & $0.0023$ & $0.0074$ & $0.0013$ & $0.0073$\\
$0.90$ & $0.15$ & $-0.0716$ & $0.0020$ & $0.0090$ & $0.0025$ & $0.0087$\\[1ex]

$0.31$ & $0.26$ & $-0.0316$ & $0.0036$ & $0.0067$ & $0.0001$ & $0.0067$\\
$0.49$ & $0.27$ & $-0.0773$ & $0.0035$ & $0.0059$ & $0.0010$ & $0.0058$\\
$0.69$ & $0.26$ & $-0.0979$ & $0.0039$ & $0.0064$ & $0.0002$ & $0.0064$\\
$0.90$ & $0.27$ & $-0.0879$ & $0.0035$ & $0.0078$ & $0.0006$ & $0.0078$\\[1ex]

$0.34$ & $0.38$ & $-0.0219$ & $0.0100$ & $0.0202$ & $0.0108$ & $0.0170$\\
$0.50$ & $0.43$ & $-0.0759$ & $0.0032$ & $0.0055$ & $0.0008$ & $0.0052$\\
$0.69$ & $0.44$ & $-0.1271$ & $0.0038$ & $0.0063$ & $0.0012$ & $0.0059$\\
$0.89$ & $0.43$ & $-0.1182$ & $0.0054$ & $0.0100$ & $0.0002$ & $0.0098$\\[1ex]

$0.54$ & $0.58$ & $-0.1211$ & $0.0108$ & $0.0160$ & $0.0016$ & $0.0159$\\
$0.70$ & $0.64$ & $-0.1474$ & $0.0049$ & $0.0075$ & $0.0003$ & $0.0075$\\
$0.89$ & $0.67$ & $-0.1612$ & $0.0059$ & $0.0107$ & $0.0022$ & $0.0105$\\
\hline\hline
\end{tabular}
\caption{$A_{\textrm{N}}$s for forward neutron production as function of 
$\braket{x_{\textrm{F}}}$ and $\braket{p_{\textrm{T}}}$.
$A_{\textrm{N}}$ and kinematic values from Fig. \ref{fig:result} (b) are listed.}
\label{tb:backward}
\end{center}
\end{table*}

The $\pi$ and $a_1$ exchange model partially reproduces the current results, but does not
explain the $x_{\textrm{F}}$ dependence.
In Fig.~\ref{fig:result} (a), $A_{\textrm{N}}$s in $0.40 < x_{\textrm{F}} < 0.60$ and
$0.60<x_{\textrm{F}}<1.00$ are consistent in $p_{\textrm{T}}<0.3$~GeV/$c$,
but a $x_{\textrm{F}}$ dependence is observed for higher $p_{\textrm{T}}$.
In Ref.~\cite{ref:pi0_a1}, spin effects by the absorptive corrections,
which are initial/final state interactions, 
start to increase from $p_{\textrm{T}}$$\sim$$0.2$~GeV/$c$. 
However, it is also expected in that calculation 
that the absolute value of the neutron $A_{\textrm{N}}$ is
larger in $0.40 < x_{\textrm{F}} < 0.60$ than that in $0.60 < x_{\textrm{F}} < 1.00$,
which is opposite to the measurements.
Other Regge poles like $\rho$ and $a_2$ may enhance the asymmetry in
$0.60 < x_{\textrm{F}} < 1.00$
because the spin effect by the $\rho$ and $a_2$ exchange can also have a finite contribution
compared to the $\pi$ and $a_1$ exchange in the higher $x_{\textrm{F}}$ 
region~\cite{ref:Mitsuka}.
More comprehensive theoretical considerations are necessary to understand the
$x_{\textrm{F}}$ dependence in $p_{\textrm{T}} > 0.3$~GeV/$c$.
Thus far no Reggeon exchange model and absorptive corrections can explain the 
$x_{\textrm{F}}$ dependence in $p_{\textrm{T}} < 0.3$~GeV/$c$, 
therefore more precise theoretical calculations, or the inclusion of new processes
other than the above production mechanism
may be necessary to explain the present results.

\section{Summary}
\label{sec:summary}

The RHICf Collaboration installed the RHICf detector at the zero-degree
area of the STAR detector and
measured the $A_{\textrm{N}}$ for forward neutron production
in polarized $p+p$ collisions at $\sqrt{s} = 510$ GeV.
This measurement covered a wide $p_{\textrm{T}}$ range with high resolution
to better understand the production mechanism for forward neutrons.
The resulting $A_{\textrm{N}}$ increases in magnitude with
$p_{\textrm{T}}$ in the high $x_{\textrm{F}}$ range, but reaches a plateau in the low
$x_{\textrm{F}}$ range.
There are indications that the asymmetries also level off at high $x_{\textrm{F}}$, 
but the magnitude increases with increasing $p_{\textrm{T}}$ bins.
No $\sqrt{s}$ dependence was observed when the RHICf data was compared with
PHENIX.
The existing theoretical calculation based on the $\pi$ and $a_1$ exchange between
two protons reproduced only part of the data.
To understand the present results, some additional spin effects beyond 
the $\pi$ and $a_1$ exchange scenario will be necessary.

\begin{acknowledgments}

We thank the staff of the Collider-Accelerator Department at Brookhaven 
National Laboratory, the STAR Collaboration and the PHENIX Collaboration 
to support the experiment. 
We especially acknowledge the essential supports from the STAR members for
the design and the construction of the detector manipulator, installation/uninstallation,
integration of the data acquisition system, operation and management of all these 
collaborative activities.
This work was supported by
the Japan-US Science and Technology 
Cooperation Program in High Energy Physics, JSPS KAKENHI 
(Nos. JP26247037, JP18H01227, and JP21H04484), 
the joint research program of the Institute for Cosmic Ray Research (ICRR), University of 
Tokyo, the NRF grants for the Center for Extreme Nuclear Matters 
(CENuM) funded by MSIT of Korea (No. 2018R1A5A1025563),
and ``UNICT " program, University of Catania.
\end{acknowledgments}




\end{document}